\documentclass{osa-article}

\journal{osajournal}

\articletype{Research Article}

\usepackage{lineno}
\usepackage{amsmath}
\usepackage{physics}
\usepackage[version=4]{mhchem}
\usepackage{siunitx}
\usepackage{graphicx}
\usepackage{svg}

\begin{document}

\title{Heterodyne coherent detection of phase modulation in a mid-infrared unipolar device}

\author{Hamza Dely,\authormark{1,*} Baptiste Chomet\authormark{1}, Thomas Bonazzi\authormark{1}, Djamal Gacemi\authormark{1}, Angela Vasanelli\authormark{1}, Axel Evirgen\authormark{2}, Olivier Lopez\authormark{3}, Benoît Darquié\authormark{3}, Filippos Kapsalidis\authormark{4}, Jérôme Faist\authormark{4}, and Carlo Sirtori\authormark{1}}

\address{
\authormark{1}Laboratoire de Physique de l'ENS, Département de physique, Ecole normale supérieure, Université PSL, Sorbonne Université, Université Paris Cité, CNRS, 75005 Paris, France\\
\authormark{2}III-V Lab, Palaiseau, 91120 Palaiseau, France\\
\authormark{3}Laboratoire de Physique des Lasers, Université Sorbonne Paris Nord, CNRS, Villetaneuse, France\\
\authormark{4}Institute of Quantum Electronics, ETH Zürich, Zürich 8093, Switzerland
}

\email{
\authormark{*}hamza.dely@ens.fr
} 



\begin{abstract}
Phase modulation is demonstrated in a quantum Stark effect modulator designed to operate in the mid-infrared at wavelength around \SI{10}{\um}. Both phase and amplitude modulation are simultaneously resolved through the measurement of the heterodyne signal arising from the beating of a quantum cascade laser with a highly stabilized frequency comb. The highest measured phase shift is more than 5 degrees with an associated intensity modulation of \SI{5}{\%}. The experimental results are in full agreement with our model in which the complex susceptibility is precisely described considering the linear voltage dependent Stark shift of the optical resonance.
\end{abstract}

\section{Introduction}\label{sec:intro}

The mid-infrared (MIR) wavelengths are considered for several applications ranging from spectroscopy to defense. Recently, it has been shown that $\lambda=\SI{10}{\um}$, in one of the atmospheric transparency windows, is a very promising candidate for new high-speed free-space optical data links thanks to its robustness to Mie scattering and long-range capabilities \cite{majumdar_optical_2005,sauvage_study_2019}. The development of efficient, powerful and room-temperature sources in this part of the electromagnetic spectrum in the last decades has stimulated research at the interface between MIR optics and telecommunications \cite{delga_free-space_2019, martini_high-speed_2001}.

Many efforts have led to the conception of semiconductor mid-infrared modulators based on different technologies such as SiGe/Si waveguides \cite{montesinos-ballester_mid-infrared_2022} or strong light-matter coupling regime in intersubband transitions \cite{pirotta_fast_2021}. Despite higher performances in terms of bandwidth compared to acousto-optic modulators, demonstrations have been limited to less than \SI{1}{\GHz} of \SI{-3}{\dB} bandwidth. In the perspective of future MIR fiber communcations, gas-filled hollow-core fibre modulators have been developped relying on photothermal effect to modulate the carrier by means of a simultaneously injected control signal \cite{jiang_broadband_2023}. Pure phase modulation has also been demonstrated in graphene-based MIR modulators  \cite{sherrott_experimental_2017,yamaguchi_low-loss_2018}, at the expense of gate biases in the order of a hundred of volts. However, these technologies may be of limited use when it comes to high-speed operation, due to either thermal inertia, or to maximum voltage limitation of high-speed generators. Recent work on unipolar quantum devices has shed light on a novel high-speed external electro-absorption amplitude modulator in the mid-infrared domain relying on quantum confined Stark effect. With several \unit{\GHz} of bandwidth, such devices are a promising solution for transmitting more than \SI{10}{\giga\bit\per\second} at \SI{9}{\um} \cite{dely_10_2022} and may write signals on the carrier even faster when embedded in a metamaterial architecture \cite{hakl_ultrafast_2021}.

Yet, telecom technologies largely exploit electro-optic phase modulators that offer a better flexibility as they can be used either as-is for phase modulation or integrated in Mach-Zehnder interferometers to provide intensity modulation \cite{pollock_waveguide_2003,ho_digital_2005,thomson_high_2013}. In addition, modern communication coding makes extensive use of simultaneous phase and amplitude modulation to multiply link datarates. In the past, some phase modulation measurements have been performed using Michelson interferometers \cite{dupont_phase_1993, dupont_mid-infrared_1993} or short mid-infrared pulses \cite{balagula_phase_2017} on other types of multiple quantum wells mid-infrared modulators.

In this work, we demonstrate that unipolar Stark-effect external modulator offers a considerable phase modulation associated to the already demonstrated amplitude modulation. Moreover, we illustrate that these two modulation regimes can be, to some extent, decoupled to select the device operation mode. A heterodyne setup is used for the coherent optical detection of amplitude and phase quadratures and their demodulation, in conjunction with a high-speed lock-in amplifier.

\section{Device modelling}\label{sec:modelling}

\begin{figure}[h]
    \centering
    \includegraphics[width=\columnwidth]{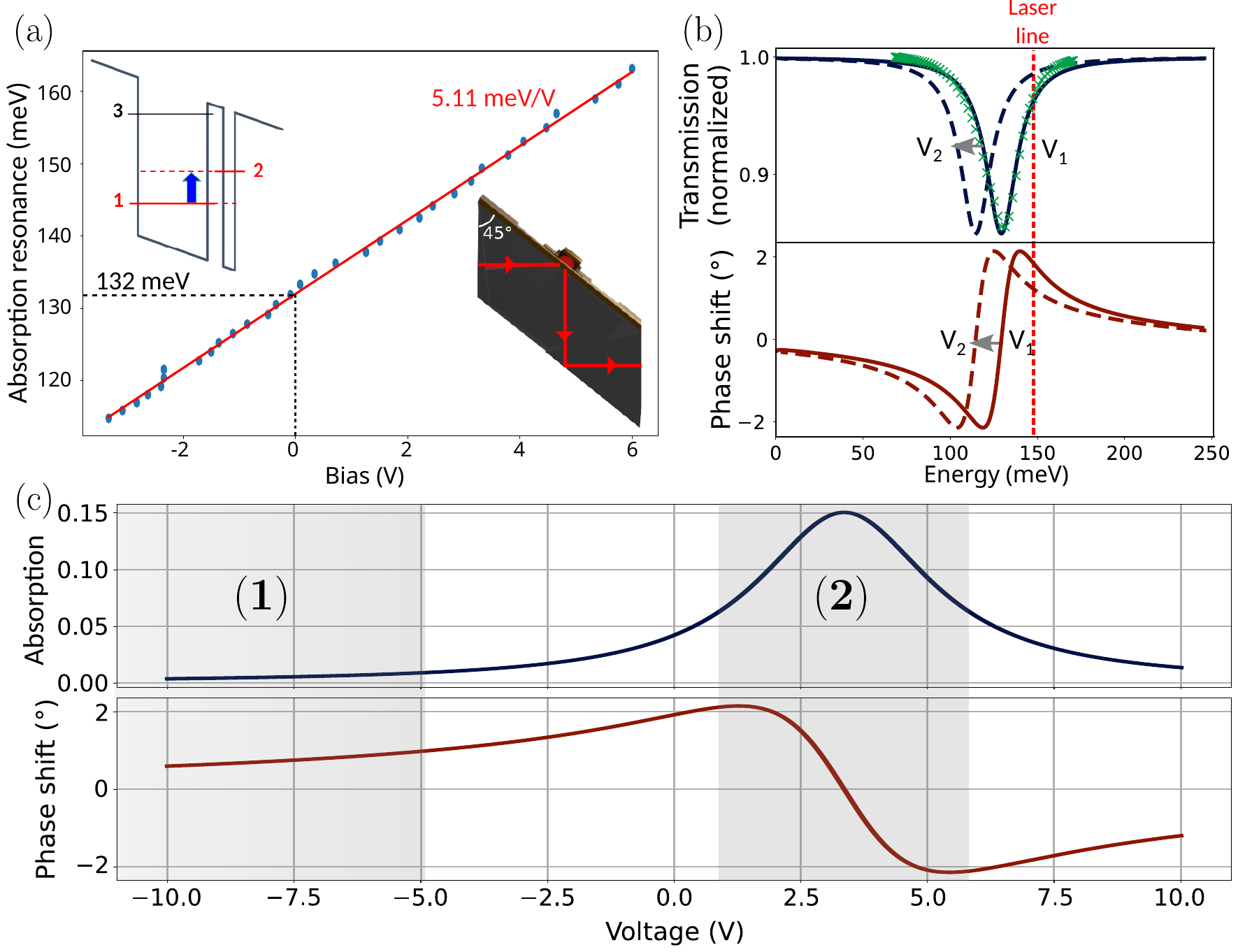}
    \caption{\label{fig:absorption_phase_vs_energy}
    (a) Absorption peak energy evolution with voltage bias for transition 1-2 shown in the upper left inset. The device is presented in the lower right inset. A linear fit of the experimental data is obtained using a tunable laser and gives a Stark tuning coefficient of \SI{5.11}{\meV\per\V}.
    (b) Experimental optical transmission of the modulator measured with a Fourier-Transform Infrared interferometer (green markers) fitted with a Lorentzian profile (dark blue line) and processed into phase variation using Kramers-Krönig relations (red curve). For different biases $V_1$ and $V_2$, the absorption peak energy shifts due to the Stark effect and so does the phase shift profile.
    (c) Evolution of the absorption and phase shift with respect to the applied bias calculated from experimental data at \SI{147}{\meV}. Regions of interest for phase modulation are highlighted in gray.}
\end{figure}

The device is a \SI{150}{\um} diameter mesa modulator based on $N_{qw}=20$ periods of a system of asymmetric coupled quantum wells \cite{dely_10_2022}. The first and second InGaAs well are \SIlist{54;26}{\angstrom} respectively, and separated by a \SI{14}{\angstrom} InAlAs barrier. Two consecutive periods are separated by \SI{200}{\angstrom} InAlAs barriers. The optical beam is coupled into the mesa through the substrate, that has a facet polished at $\theta=\SI{45}{\degree}$ (Fig. \ref{fig:absorption_phase_vs_energy}(a)). The interaction length with the active region of the modulator is $L=2N_{qw}L_{qw}\sin^2\theta/\cos\theta=\SI{100}{\nm}$ and a peak absorption of \SI{15}{\percent} is observed at \SI{132}{\meV} (Fig. \ref{fig:absorption_phase_vs_energy}(b)). The absorption data measured at \SI{0}{\V} (green markers) is fitted with a Lorentzian profile (dark blue curve) and the corresponding phase evolution is obtained with the help of Kramers-Krönig relations (red curve), with a maximum phase shift in excess of $\pm\SI{2}{\degree}$. Under bias, the absorption peak shifts linearly with the applied voltage as represented in Fig. \ref{fig:absorption_phase_vs_energy} (a). The Stark tuning coefficient, $m=\SI{5.11}{\meV\per\V}$, has been measured with a tunable MIR laser (Daylight Solutions MIRcat-QT mid-IR). The device operation, illustrated in Fig. \ref{fig:absorption_phase_vs_energy}(b), can be figured as an absorption peak, and the associated phase variation, that move backward and forward in energy according to the bias. As a consequence, we can infer the absorption and phase shift for a fixed incident wavelength as a function of the voltage by mapping energies in the absorption spectra to voltages using the Stark coefficient $m$ as reported on Fig. \ref{fig:absorption_phase_vs_energy}(c). This assumes that the oscillator strength as well as the linewidth of the transition are invariant with respect to the bias which will be justified \textit{a posteriori}.
Two regions of interest for phase modulation are highlighted: the first area from \SI{-10}{\V} to \SI{-5}{\V} exhibits low linear absorption and phase shift while the second one around the absorption peak provides a higher linear phase shift at the expense of a strong absorptivity.

The behavior of our unipolar device can be described considering as a first approximation a voltage-dependent electric susceptibility $\Delta\chi\left(\omega, V\right)$ at angular frequency $\omega$ and under a bias $V$ modelled by a Lorentzian function such that
\begin{equation}\label{eq:susceptibility}
    \Delta\chi(\omega, V) = \frac{\omega_p^2 f}{\omega}\frac{i\gamma + \left(\omega - \omega_0 - mV/\hbar\right)}{\gamma^2 + \left(\omega - \omega_0 - mV/\hbar\right)^2}
\end{equation}
where $\omega_0$ is the angular frequency corresponding to the intersubband transition responsible for the absorption and $f$ its oscillator strength, $\omega_p$ the plasma frequency, $\gamma$ the half-width at half-maximum of the intersubband transition linewidth. Equation \ref{eq:susceptibility} is derived from preliminary modeling work based on a density matrix approach, which clearly validates that the physical interpretation of the absorption peak moving back and forth is valid as long as the frequency of the signal is lower than the coherence time of the transition, in the order of 500 fs to 1 ps \cite{bonvalet_femtosecond_1996}. From the usual expression of a plane electromagnetic wave $E_0\exp\left(i\left(\frac{n\omega}{c}L-\omega t\right)\right)$ and the definition of the refractive index $n^2=n^2_0\left(1+\Delta\chi\left(\omega, V\right)\right)$ with $n_0\sim 3.4$ the index of the bulk semiconductor material, the absorption $\alpha$ and the phase shift $\phi$ after the modulator can be expressed using the parameters of equation \ref{eq:susceptibility} as:
\begin{equation}\label{eq:intensity_phase}
\left\{
    \begin{aligned}
        \alpha = -\frac{\omega n_0 L}{c}\Im\left(\Delta\chi\right) &= -\frac{n_0\omega_p^2 f L}{c}\frac{\gamma}{\gamma^2+\left(\omega-\omega_0-mV/\hbar\right)^2} \\
        \phi = \frac{1}{2}\frac{\omega n_0 L}{c}\Re\left(\Delta\chi\right) &= \frac{1}{2}\frac{n_0\omega_p^2 f L}{c}\frac{\omega-\omega_0-mV/\hbar}{\gamma^2+\left(\omega-\omega_0-mV/\hbar\right)^2}
    \end{aligned}
\right.
\end{equation}
In the following, we describe a heterodyne setup that allows the measurement of both phase and amplitude modulation properties of our Stark modulator.

\section{Experimental setup}\label{sec:setup}
\begin{figure}[h!]
    \centering
    \includegraphics[width=\columnwidth]{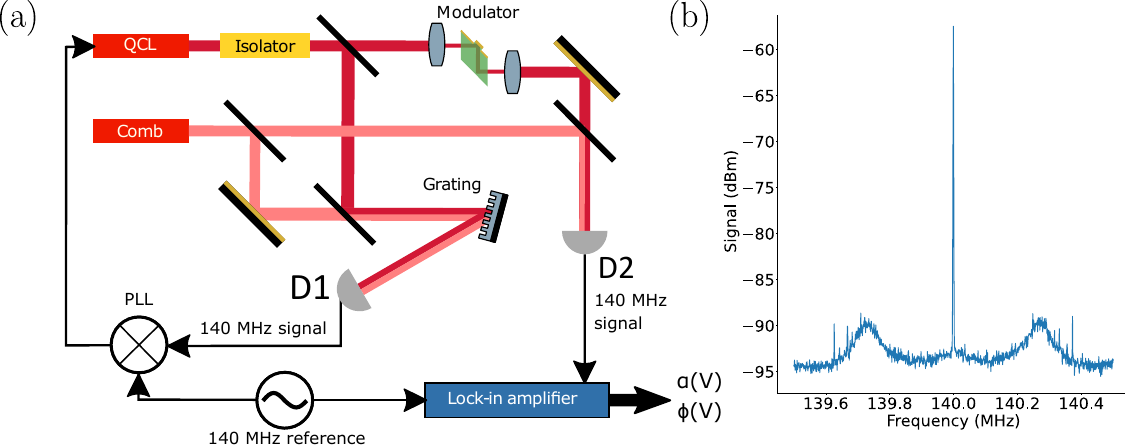}
    \caption{\label{fig:heterodyne_setup}
    (a) Heterodyne setup used to measure both amplitude and phase modulation of a coherently phase locked QCL to a tooth of a mid-IR frequency comb. A first detector D1 generates the beat-note signal between the QCL and one tooth of the comb which is subsequently phase locked to a reference signal at 140 MHz by applying a correction to the QCL’s current. An optical isolator is used to avoid feedback in the QCL. The second detector D2 generates the heterodyne signal between the modulated QCL and the same tooth of the mid-IR comb. The resulting beat signal is demodulated in a lock-in amplifier using the same 140 MHz reference signal. (b) Heterodyne beat-note at \SI{140}{\MHz} observed with a \SI{1}{\kHz} resolution bandwidth.}
\end{figure}

The experimental setup is depicted in Fig.\ref{fig:heterodyne_setup}(a). It consists of a heterodyne detection scheme where the beam of a modulated laser and the beam of a second laser are superimposed and focused onto a high speed detector. Using a phase locking process the frequency difference $\Delta\omega$ between the two lasers prior modulation is stabilized \cite{sow_widely_2014,chomet_highly_2023}. The resulting signal at the output of the detector oscillates at $\Delta\omega$ and is given by:
\begin{equation}
I_{het}\propto I_{2} + I_{1}e^{-\alpha\left(V\right)} + 2\sqrt{I_{2}I_{1}}e^{-\frac{1}{2}\alpha\left(V\right)}\cos\left(\Delta\omega t + \phi\left(V\right) + \delta\phi\right)
\label{eq:heterodyne}
\end{equation}
where $I_{1}$, $I_{2}$ are the intensities of the modulated laser and the second laser respectively. $\delta\phi$ is the overall phase difference between the two lasers (due to the difference in optical paths, modulator substrate,...). Eq.\ref{eq:heterodyne} shows that both amplitude and phase of the modulator can be simultaneously retrieved through demodulation of the signal $I_{het}$ at the frequency $\Delta\omega$. 

The two lasers of the scheme in Fig.\ref{fig:heterodyne_setup}(a) are phase locked to measure both amplitude and phase. The first source is a Fabry-Perot quantum cascade laser (QCL) from ETH Zürich, electrically injected with a low noise current driver (ppqSense QubeCL CD-10). It is typically operated at a temperature of \SI{278}{\kelvin} and delivers up to \SI{20}{\mW} in a single transverse and longitudinal mode at a wavelength of \SI{8.4}{\um} in the \qtyrange{400}{540}{\mA} current range. The second source is a commercial mid-IR comb (Menlo Systems) based on a difference frequency generation (DFG) process in a non-linear crystal between an ultra-low noise \SI{1.5}{\um} Erbium doped fiber frequency comb and its extension in a wavelength shifting fiber towards the \SI{1.8}{\um} window. The system delivers an optical beam at \SI{8.4}{\um} wavelength (\qtyrange{7.4}{9.4}{\um}) with an average output power of \SI{2.2}{\mW} at a repetition rate of \SI{100}{\MHz}. The QCL and the mid-IR comb beams are first superimposed by means of a 50/50 germanium (\ce{Ge}) beam splitter, filtered by a diffracting grating with 150 grooves/mm, and sent onto a \SI{800}{\MHz} bandwidth mercury-cadmium-telluride (MCT) detector $D_1$ (ViGO PVI-4TE-10.6). The heterodyne signal between the QCL line and the second closest tooth of the mid-IR comb at the output of the detector $D_1$ is compared to a synthesized reference signal typically generated around $\Delta\omega = 2\pi\times\SI{140}{\MHz}$ in this work. A phase-lock loop (PLL) is used to apply a correction signal directly to the QCL’s current. As shown in Fig. \ref{fig:heterodyne_setup}(b), in closed loop operation a narrow beat note linewidth (resolution bandwidth limited) of the heterodyne signal is observed with a \SI{45}{\dB} signal-to-noise ratio peak in \SI{1}{\kHz} resolution-bandwidth. 

Part of the QCL light is now sent through the modulator and recombined with the mid-IR comb onto the detector $D_2$ (similar to $D_1$) that generates the signal of Eq.\eqref{eq:heterodyne}. The latter is compared to the reference signal using a lock-in amplifier (UHFLI, Zürich Instrument) that allows to set some additional bandwidth (a few tens of kHz here) around the beatnote frequency so that we can observe the real-time evolution of both $\alpha (V)$ and $\Phi (V)$.

\section{Phase shift and amplitude measurements}\label{sec:measurements}

\begin{figure}[h]
    \centering
    \includegraphics[width=\columnwidth]{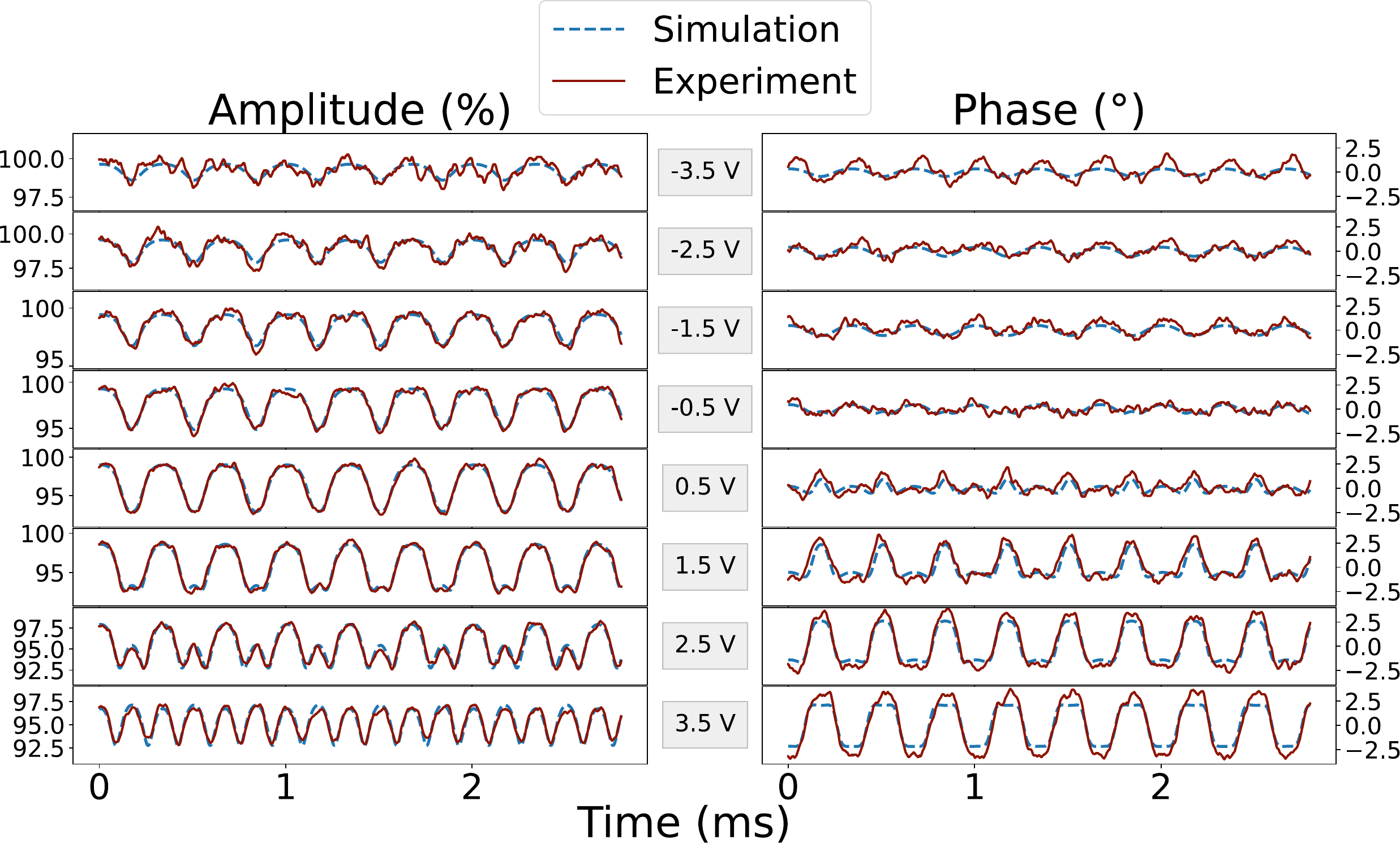}
    \caption{\label{fig:amplitude_phase_DC+5Vpp}
    Simulation (blue) and experimental data (red) for amplitude and phase shifts of a laser at \SI{8.4}{\um} wavelength. AC bias is kept at \SI{5}{\volt} peak-to-peak while DC bias is swept from \SIrange{-3.5}{3.5}{\volt}.}
\end{figure}

Fig. \ref{fig:amplitude_phase_DC+5Vpp} presents the output from the two channels of the lock-in, phase and amplitude, as a function of time (red curves) and their respective simulations based on Eq. \eqref{eq:intensity_phase} (blue curves) for different harmonic voltage waveforms $V=V_{DC}+\delta V\cos\left(\Omega t\right)$ applied on the modulator. The angular frequency is kept constant at $\Omega=2\pi\times\SI{3}{\kHz}$ as well as the voltage amplitude $\delta V=\SI{2.5}{\V}$ while DC bias $V_{DC}$ ranges from \SIrange{-3.5}{3.5}{\volt}. At $V_{DC}=\SI{-3.5}{\volt}$, the modulator operates in the tail of its absorptivity (see Fig. \ref{fig:absorption_phase_vs_energy}c), leading to low values of both amplitude and phase modulation. The optimum working point for amplitude modulation is achieved on the side of the absorption peak at a DC bias of \SI{0.5}{\volt}. Here the phase modulation is negligible end exhibits a boost in its anharmonicity, as it occurs on one of the extrema of the phase variation. Increasing further DC bias leads to a decrease of amplitude modulation, as we shift the working point on the absorption peak. The frequency doubling appears now on amplitude modulation while phase modulation is maximum and shows a close to \SI{5}{\degree} peak-to-peak shift.

Notice that due to a low signal-to-noise ratio, the signal needs to be averaged on 256 acquisitions. These measurements are in excellent agreement with the simulations. The signal-to-noise ratio could be improved with a substantial increase of the optical power, which was not possible in our experiment because the two detectors saturate at very low power. In addition, using two overlapping distributed feedback lasers instead of an optical comb and a grating could yield some improvements in terms of signal to noise ratio as all unnecessary optical frequencies contributing to the detector saturation would be discarded. The phase-shift effect could be increased by choosing a waveguide geometry for the modulator with ridges of a few hundreds of micrometers long, improving the signal-to-noise ratio.

\section{Simulations for waveguide modulator}

\begin{figure}[h!]
    \centering
    \includegraphics[width=\columnwidth]{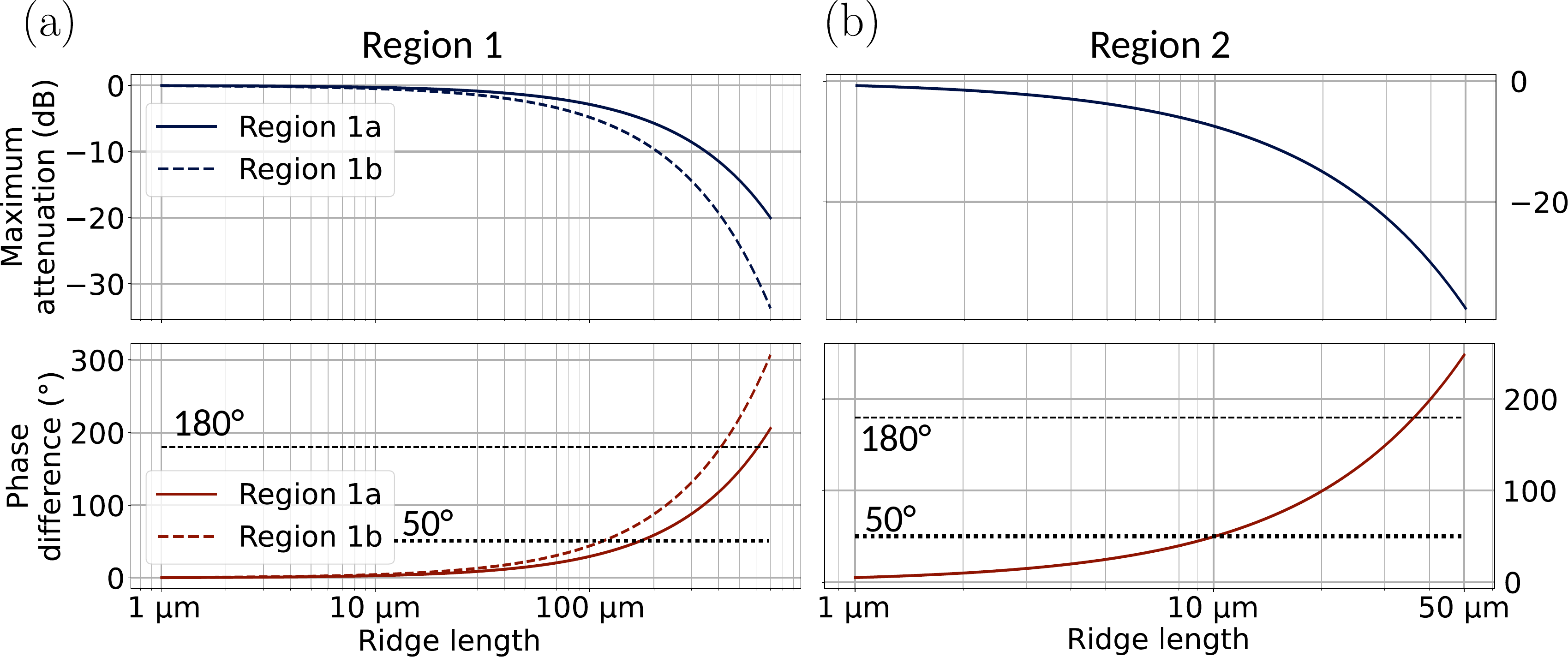}
    \caption{\label{fig:alpha}
    Maximum attenuation and phase difference obtained for waveguide modulators in the two different modulation regions highlighted in fig. \ref{fig:absorption_phase_vs_energy}c for different modulator ridge lengths: (a) Region 1 with its two sub-regions: 1a (\qtyrange{-12.5}{-7.5}{\V}) and 1b (\qtyrange{-10}{-5}{\V}) and (b) Region 2 (\qtyrange{1}{6}{\V}).}
\end{figure}
Our study is based on a $\SI{1}{\um}$ thick mesa device producing a \SI{4}{\degree} phase shift, but the model introduced in section \ref{sec:modelling} is useful to estimate the potential of future mid-infrared waveguide modulators based on a the same active region (AR) to avoid the need of excessive modulating voltages on the device. In a waveguide geometry, the coupling between the active region and the gaussian transverse electromagnetic mode $E=E_0\exp\left(-r^2/w^2\right)$, with $w$ the beam waist, propagating in the waveguide needs to be accounted for through an overlap factor $\xi$ in the growth direction defined as \cite{paiella_quantum_2010}
\begin{equation}
    \xi=\frac{\displaystyle\int_{AR}E^2\,dz}{\displaystyle\int^\infty_{-\infty}E^2\,dz}
\end{equation}
The value of the overlap factor $\xi$, for the fundamental mode of a ridge waveguide of height \SI{10}{\um} with a \SI{300}{nm} AR situated at the core of the waveguide, can be estimated at around \SI{10}{\%}. Figure \ref{fig:alpha} shows the evolution of the maximum intensity absorbed and phase difference induced in a ridge modulator as a function of its length for the two dashed regions in Fig. \ref{fig:absorption_phase_vs_energy}(c). In the first sub-region (1a, \SIrange{-12.5}{-7.5}{\V}) a phase difference of $\pi$ is obtained for $V_\pi=\SI{5}{\V}$ for a ridge length of \SI{600}{\um} at the cost of over \SI{18}{\dB} of optical power attenuation. When shifting the voltage limits towards the absorption peak (1b, \qtyrange{-10}{-5}{\V}), the required length is slightly decreased at \SI{400}{\um} with similar power losses. Such lengthy ridges can introduce some difficulties when operated at very high frequencies in the order of \SI{100}{\GHz} with vacuum wavelengths in the millimeter range or below. In that case, similar phase shift performances could be achieved in the peak absorption region (2, \SIrange{1}{6}{\V}) with a ten times shorter ridge ($L=\SI{35}{\um}$) but at the cost of a higher \SI{28}{\dB} signal attenuation (Fig. \ref{fig:alpha}b).

\section{Conclusion}\label{sec:conclusion}

Voltage controlled phase modulation has been achieved in a mid-infrared unipolar quantum optoelectronic device that relies on the Stark effect. The phase modulation is coupled to that of amplitude following Kramers-Krönig relations. Remarkably in our device the perfect linear dependence of the Stark shift on the applied bias consents to express Kramers-Krönig relations as a function of the voltage. A simple Lorentzian function, describing the amplitude modulation, with a linear voltage dependant resonance models precisely the behaviour of the complex susceptibility. Selecting carefully the voltage range applied to the modulator, the phase can be modified without significant effects on the amplitude and vice-versa. The modulator can furnish a phase shift of 5° with only few percent losses on amplitude. This would be enough to encode extra information, thus increasing the number of channels of information for free space communications. In addition, the Stark modulator active region would enable the design of high-speed MIR electro-optic modulators with waveguides of a few tens of micrometers bringing up to 180° phase shift within a \SI{5}{\V} voltage range at the expense of a ~\SI{20}{\dB} attenuation.

Our results are a first step towards quadrature amplitude modulation which is the standard protocol for data transmission in telecom technology. By cascading, for instance, two modulators, more complex and industry-standard keying than simplistic NRZ patterns will be enabled. A heterodyne setup with a wide band detector can, therefore, be used to recover the phase information by down-converting it from the optical to any sub-THz domain, including radiofrequencies, microwaves or mm-wave. Existing hardware at these frequencies (Wi-Fi, cellular, radio networks) could efficiently relay and/or process these signals. Moreover, telecom application would allow to remove the servo-loop by the use of adapted coding scheme and phase recovery algorithms, at the expense of an increased latency on the channel \cite{ip_feedforward_2007,navarro_carrier_2016}. Finally, phase modulation will strengthen also the ability to perform coherent detection, which is at the moment the most sensitive approach to detect faint signals in the mid-infrared range.

\begin{backmatter}
\bmsection{Funding}
This work has been supported by ENS-Thales Chair, Agence Nationale de la Recherche projects LIGNEDEMIR (ANR-18-CE09-0035) and CORALI (ANR-20-CE04-0006), H2020 Future and Emerging Technologies (Project cFLOW, Project Qombs), CNRS Renatech network, and the Region Ile-de-France in the framework of DIM SIRTEQ.

\bmsection{Disclosures}
The authors declare no conflicts of interest.

\bmsection{Data availability} Data underlying the results presented in this paper are not publicly available at this time but may be obtained from the authors upon reasonable request.

\end{backmatter}

\bibliography{sample}

\end{document}